\documentclass{elsart}
\usepackage{color,setspace,times,delarray}
\usepackage{amssymb,amsmath,graphicx,rotating,subfigure,url}
\usepackage{lineno}
\usepackage[notablist,nofiglist]{endfloat}

\journal{Physica A} 

\begin{document}

\begin{frontmatter}
\title{Nonlinear behavior of the Chinese SSEC index with a unit root: Evidence from threshold unit root tests}
\author[SS,CES]{Xi-Yuan Qian},
\ead{xyqian@ecust.edu.cn}
\author[CES,BS]{Fu-Tie Song},
\ead{ftsong@ecust.edu.cn}
\author[SS,CES,BS,RCSE]{Wei-Xing Zhou\corauthref{cor}}
\corauth[cor]{Corresponding author. Address: 130 Meilong Road, P. O.
Box 114, School of Business, East China University of Science and
Technology, Shanghai 200237, China, Phone: +86 21 64253634, Fax: +86
21 64253152.} \ead{wxzhou@ecust.edu.cn}


\address[SS]{School of Science, East China University of Science and Technology, Shanghai 200237, China}
\address[CES]{Center for Econophysics Studies, East China University of Science and Technology, Shanghai 200237, China}
\address[BS]{School of Business, East China University of Science and Technology, Shanghai 200237, China}
\address[RCSE]{Research Center of Systems Engineering, East China University of Science and Technology, Shanghai 200237, China}

\begin{abstract}
We investigate the behavior of the Shanghai Stock Exchange Composite
(SSEC) index for the period from 1990:12 to 2007:06 using an
unconstrained two-regime threshold autoregressive (TAR) model with
an unit root developed by Caner and Hansen. The method allows us to
simultaneously consider non-stationarity and nonlinearity in
financial time series. Our finding indicates that the Shanghai stock
market exhibits nonlinear behavior with two regimes and has unit
roots in both regimes. The important implications of the threshold
effect in stock markets are also discussed.
\end{abstract}

\begin{keyword}
Threshold autoregressive (TAR) model; Unit root; Chinese stock
market; Regime change; Crashes
\end{keyword}

\end{frontmatter}

\section{Introduction}
\label{s1:Intro}

In the past three decades since the Third Plenary Session of the
11th Central Committee of the Communist Party of China in December
1978, China has paved a gradual transition from a centrally planned
economy to a market economy and the economy has experienced
unprecedented growth. During this period, one of the most important
developments has been the reopening and operation of the Chinese
stock market. Before the foundation of People's Republic of China,
the Shanghai Stock Exchange was the third largest worldwide (after
New York and London Stock Exchanges) and had remarkable influence on
other world-class financial markets \cite{Su-2003}. After 1949,
China implemented policies of a socialist planned economy and the
government controlled entirely all investment channels. In 1981, the
central government began to issue treasury bonds to raise capital to
cover its financial deficit, which reopened China's securities
markets. The first market for government-approved securities was
founded in Shanghai on November 26, 1990 and started operating on
December 19 of the same year under the name of Shanghai Stock
Exchange (SHSE). Shortly after, the Shenzhen Stock Exchange (SZSE)
was established on December 1, 1990 and started its operations on
July 3, 1991. The size of the Chinese stock market has increased
remarkably \cite{Zhou-Sornette-2004a-PA}. There are increasing
interests in the academic studies of the Chinese stock market.

It is well-known that most econometric models are constructed based
on the assumption that the variables are stationary. It is thus very
important to perform unit root test for stationarity. Empirical
studies on the US markets find mixed results concerning the presence
of a unit root in the behavior of stock indexes
\cite{Narayan-2006-MCS}. The situation seems alike in the Chinese
stock market. Xie, Gao and Ma applied the Phillips-Perron (PP) test
to the weekly data of the Shanghai Stock Exchange Composite (SSEC)
index from 12/21/1990 to 03/02/2001 and found that the null
hypothesis that the logarithm of the index has a unit root cannot be
rejected at the significance level of $5\%$
\cite{Xie-Gao-Ma-2002-cnJQTE}. Cheng, Wu and Zhou tested the unit
root property in the daily data of SSEC from 01/02/1998 to
12/31/2001 using the augmented Dickey-Fuller (ADF) approach and
reached a similar conclusion \cite{Cheng-Wu-Zhou-2003-cnCJMS}. In
contrast, Dai, Yang and Zhang investigated the daily data of the
SSEC index from 12/19/1990 to 06/18/2004 with the ADF approach and
found that the unit root null is rejected at the $5\%$ level of
significance \cite{Dai-Yang-Zhang-2005-cnSE}.

On the other hand, numerous evidence shows that there exists
threshold nonlinearity in the behavior of stocks
\cite{Shively-2003-QREF,Chen-So-Gerlach-2005-ANZJS}. It is thus
helpful to distinguish non-stationarity from nonlinearity in the
stock market behavior. This task can be done by adopting the
threshold unit root test developed by Caner and Hansen
\cite{Caner-Hansen-2001-Em}. Briefly speaking, Caner and Hansen use
threshold autoregressive (TAR) model to test for a threshold effect
and then perform unit root tests on both regimes if exist. Recently,
Narayan applied this method to the stock prices of Australia (ASX
All Ordinaries, monthly data from 1960:01 to 2003:04) and New
Zealand (NZSE Capital Index, monthly data from 1967:01 to 2003:04)
\cite{Narayan-2005-AE}, and the US stock price index (NYSE common
stocks, monthly data from 1964:06 to 2003:04)
\cite{Narayan-2006-MCS}. The main finding is that the stock prices
in the three markets are generated by nonlinear processes and can be
characterized by unit root processes.

This work attempts to add to the existing literature by testing for
nonlinearity and unit root property of the Chinese stock price index
SSEC, which is monthly over the period from 1990:12 to 2007:06. We
find that the monthly SSEC index behaves nonlinearly with a unit
root in both regimes. The paper is organized as follows. Section
\ref{s1:method} reviews briefly the procedure of the threshold unit
root test of Caner and Hansen \cite{Caner-Hansen-2001-Em}. Section
\ref{s1:analysis} presents the empirical results. And Section
\ref{s1:conclusion} gives some conclusive remarks.

\section{Econometric methodology}
\label{s1:method}

In this section, we describe briefly the econometric methodology of
the threshold unit root tests proposed by Caner and Hansen
\cite{Caner-Hansen-2001-Em}. A rigorous presentation with
assumptions, theorems and proofs can be found in their seminal paper
\cite{Caner-Hansen-2001-Em}. See also references
\cite{Basci-Caner-2005-SNDE,Basci-Caner-Yoon-2006-SNDE} for a
tutorial example.

\subsection{Model specification and calibration}

Following the work of Caner and Hansen \cite{Caner-Hansen-2001-Em},
we adopt a two-regime threshold autoregression (TAR) model with an
order of $k$. The mathematical expression of the TAR($k$) model
reads
\begin{equation}
\label{Eq:TAR}
 \Delta{y_t}={\theta_{1}x_{t-1}I(Z_{t-1}<\lambda)
            + \theta_{2}x_{t-1}I(Z_{t-1}\geqslant\lambda)+\epsilon_t}
\end{equation}
with
\begin{equation}
\label{Eq:x}
 x_{t-1}=(y_{t-1}, 1,\Delta{y_{t-1}}, \cdots,
 \Delta{y_{t-k}})^{'}~,
\end{equation}
where $y$ is the logarithm of the SSEC index for $t=1,2,\cdots,T$,
$\epsilon_t$ is an i.i.d. error, $I({\rm{expression}})$ is the
indicator function that equals to 1 if the expression in the
parentheses is true and 0 otherwise, $Z_{t}=y_{t}-y_{t-m}$ for some
$m \geqslant 1$ is the threshold variable, and $k \geqslant 1$ is
the autoregressive order. The variable $Z_t$ has clear financial
meaning acting as {\em{return}} at the time horizon of $m$ months.
The threshold parameter $\lambda$ is unknown and represents the
level of the variable $y_t$ that triggers a ``regime change'', if
any. The components of $\theta_1$ and $\theta_2$ can be partitioned
as follows:
\begin{equation}
 \left\{
 \begin{array}{ccc}
   \theta_1&=&\left(\rho_1, \beta_1,\alpha_1\right)\\
   \theta_2&=&\left(\rho_2, \beta_2, \alpha_2\right)
 \end{array}
 \right.,
 \label{Eq:theta}
\end{equation}
where $\rho_1$ and $\rho_2$ are slope coefficients on $y_{t-1}$,
$\beta_1$ and $\beta_2$ are scalar intercepts, and $\alpha_1$ and
$\alpha_2$ are $1 \times k$ vectors containing the slope
coefficients on dynamics regressors $(\Delta y_{t-1}, \ldots, \Delta
y_{t-k})$ in the two regimes.

In order to calibrate model (\ref{Eq:TAR}), the concentrated least
squares approach is usually utilized. The regression procedure is
carried out for each value of $m$. The value of $\lambda$ is taken
from a compact interval $[\lambda_1,\lambda_2]$ in which $\lambda_1$
and $\lambda_2$ are determined by the following constraints
\begin{equation}
 \left\{
 \begin{array}{ccc}
   \Pr(Z_t\leqslant\lambda_1)&=&\pi_1\\
   \Pr(Z_t\leqslant\lambda_2)&=&\pi_2
 \end{array}
 \right.,
 \label{Eq:lambda}
\end{equation}
where $0<\pi_1<\pi_2<1$ and $\pi_1+\pi_2=1$. In this work, we impose
$\pi_1=0.15$ \cite{Basci-Caner-2005-SNDE}. For each
$\lambda\in[\lambda_1,\lambda_2]$, the parameters $\rho$'s,
$\beta$'s and $\alpha$'s are estimated by minimizing the objective
function
\begin{equation}
 Q(\lambda,m) = \sum_{t=1}^T\epsilon_t(\lambda,m)^2~.
 \label{Eq:Q}
\end{equation}
Let $\hat{\epsilon}_t(\lambda,m)$ represents the residual from the
ordinary least squares for given $\lambda$ and $m$. Then the least
squares estimate $\hat\lambda$ of the threshold parameter is given
by
\begin{equation}
 \hat\lambda = \min_{\lambda\in[\lambda_1,\lambda_2]} \hat{Q}(\lambda,m)~.
 \label{Eq:lambda:hat}
\end{equation}
Note that $\hat\lambda$ and other estimates of parameters are
dependent of $m$.

\subsection{Test for threshold effect}

In model (\ref{Eq:TAR}), a question of particular interest is
whether or not there is a threshold effect. The threshold effect
disappears under the null hypothesis
\begin{equation}
 H_0: \theta_{1}=\theta_{2}~,
 \label{Eq:H0}
\end{equation}
which is tested using a standard heteroskedastic-consistent Wald
test \cite{Caner-Hansen-2001-Em}. The Wald statistic is
\begin{equation}
 W=W(\hat\lambda)=\sup_{\lambda \in [\lambda_1,\lambda_2]} W(\lambda)~.
 \label{Eq:W}
\end{equation}
If the null hypothesis can not be rejected, there is no threshold
effect, in which case the two vectors of coefficients are identical
between the two regimes ($\theta_1=\theta_2$). Caner and Hansen also
show that $\sup_{\lambda \in [\lambda_1,\lambda_2]} W(\lambda)$ has
a non-standard asymptotic null distribution and propose a bootstrap
method to compute the asymptotic critical values and $p$-values
\cite{Caner-Hansen-2001-Em}.

\subsection{Tests for threshold unit root}

When there are two regimes delimited by a threshold, we have two
parameters $\rho_1$ and $\rho_2$ controlling the stationarity of the
process $y_t$. The null hypothesis is
\begin{equation}
 H_0: \rho_1=\rho_2=0~.
 \label{Eq:TUR:H0}
\end{equation}
When the null hypothesis $H_0$ holds, the process $y_t$ has a unit
root and model (\ref{Eq:TAR}) can be expressed in terms of the
stationary difference $\Delta{y_t}$. An alternative hypothesis to
the null $H_0$ is
\begin{equation}
 H_1: \rho_1<0~~~~{\rm{and}}~~~~\rho_2<0~.
 \label{Eq:TUR:H1}
\end{equation}
When $H_1$ holds, the process $y_t$ is stationary and ergodic in
both regimes \cite{Caner-Hansen-2001-Em}. Another alternative deals
with a partial unit root, which is expressed as follows
\begin{equation}
 H_2: \left\{
 \begin{array}{cccc}
 \rho_1<0&{\rm{and}}&\rho_2=0, & {\rm{or}}\\
 \rho_1=0&{\rm{and}}&\rho_2<0.  &
 \end{array}
 \right.
 \label{Eq:TUR:H2}
\end{equation}
When $H_2$ holds, the process $y_t$ have a unit root in one regime
and is stationary in the other showing mean reversion behavior.

To test the null hypothesis $H_0$ against its two alternatives $H_1$
and $H_2$, there are two Wald tests that apply. The statistic of
one-sided Wald test against the unrestricted alternative
$\rho_1\neq0$ or $\rho_2\neq0$ is
\begin{equation}
 R_1 = t_1^2I(\hat\rho_1<0) + t_2^2I(\hat\rho_2<0)~,
 \label{Eq:R1}
\end{equation}
while that of the two-sided Wald test against $\rho_1<0$ or
$\rho_2<0$ is
\begin{equation}
 R_2 = t_1^2 + t_2^2~,
 \label{Eq:R2}
\end{equation}
where $t_1$ and $t_2$ are the $t$ ratios for $\hat\rho_1$ and
$\hat\rho_2$. We note that this term $t$ should not be confused with
the time $t$ in Eq.~(\ref{Eq:TAR}). In order to further discriminate
the stationarity in the two regimes, we can examine the negative of
the $t$ statistics $-t_1$ and $-t_2$.

\section{Application to the SSEC index}
\label{s1:analysis}

\subsection{The data set}

We analyze the whole time series of the SSEC index. The components
of the SSEC index consist of all stocks listed on the Shanghai Stock
Exchange, including both A shares and B shares. It is thus an
overall index reflecting the price fluctuations of the overall
Shanghai stock market. The index was officially released since July
15, 1991, tracing back to December of 1990. Monthly data over the
period from 1990:12 to 2007:06 are utilized for analysis.
Specifically, we retrieve the closing prices of the last trading
days of all months and calculate the logarithms, which give the time
series $y_t$ defined in the preceding section.

\subsection{Unit root tests}

As a first step we perform conventional unit root tests of the
monthly SSEC index without taking into account possible
nonlinearity. The augmented Dickey-Fuller (ADF)
\cite{Dickey-Fuller-1981-Em}, Phillips-Perron (PP)
\cite{Phillips-Perron-1988-Bm}, and
Kwiatkowski-Phillips-Schmidt-Shin (KPSS)
\cite{Kwiatkowski-Phillips-Schmidt-Shin-1992-JEm} tests are adopted.
We test for unit root in both the logarithm of SSEC $y_t$ and its
first-order difference $\Delta{y_t}$. For the ADF and PP tests, the
null hypothesis is that $y_t$ (resp. $\Delta{y_t}$) has a unit root,
which utilizes the $t$-statistic. In contrast, the null of the KPSS
method is the stationarity of the variable and uses the
LM-statistic. The results are presented in Table \ref{TB:UnitRoot}.
All tests indicate that the SSEC index is a unit root process while
its first-order difference is stationary.

\begin{table}[htbp]
\centering
    \caption{Unit root tests of the monthly SSEC index without threshold effect.
          The augmented Dickey-Fuller (ADF), Phillips-Perron (PP) and Kwiatkowski-Phillips-Schmidt-Shin (KPSS)
          tests are adopted. The left panel tests for unit root in the logarithm of SSEC $y_t$,
          while the right panel tests for unit root in the first-order difference.
          $V_{x\%}$ is the critical value at the $x\%$ significance level.}
    \medskip
    \label{TB:UnitRoot}
    \begin{tabular}{cccccccccccccccccc}
    \hline\hline
     & \multicolumn{5}{@{\extracolsep\fill}c}{$\log {\rm{SSEC}}$, $y_t$} &
    & \multicolumn{5}{@{\extracolsep\fill}c}{$\Delta\log {\rm{SSEC}}$, $\Delta{y_t}$}  \\
    \cline{2-6} \cline{8-12}
        method & statistic & $V_{10\%}$ & $V_{5\%}$ & $V_{1\%}$ & $p$-value & %
       & statistic & $V_{10\%}$ & $V_{5\%}$ & $V_{1\%}$ & $p$-value \\\hline %
   ADF & -0.76 & -1.62 & -1.94 & -2.58 & 0.877 && -2.56 & -1.62 & -1.94 & -2.58 & 0.011 \\%
   PP  &  1.55 & -1.62 & -1.94 & -2.58 & 0.970 && -14.6 & -1.62 & -1.94 & -2.58 & 0.000 \\%
   KPSS&  1.28 &  0.35 &  0.46 &  0.73 & 0.000 &&  0.16 &  0.35 &  0.46 &  0.73 & $\gg$0.1   \\%
\hline\hline
    \end{tabular}
\end{table}

\subsection{Threshold effect}

We now use the Wald test to examine whether we can reject the linear
autoregressive model in favor of a threshold model. In our model, we
adopt that $k=12$. In Table \ref{TB:LinearityTest}, we report the
results of the Wald test. Also listed are the bootstrap critical
values at three conventional levels $10\%$, $5\%$, and $1\%$ and the
bootstrap $p$-values for threshold variables of the form
$Z_t=y_{t}-y_{t-m}$ for different delay parameters $m$ ranging from
1 to 12. The bootstrapping is carried out with 1000 and 2000
replications. The results are qualitatively the same for both cases
so that we report below the results with 2000 replications. For all
$m$, the null hypothesis $\theta_1=\theta_2$ of linearity is
rejected at the significance level of $1\%$. In other words, the
presence of a threshold effect in the monthly SSEC index is
statistically significant with a $99\%$ confidence level. According
to these results, the linear AR model can be rejected in favor of
the TAR model.


\begin{table}[htp]
  \centering
  \caption{Wald tests for a threshold effect in the monthly SSEC index for different lags $m$.
          The second row gives the Walt-statistics $W$ for different $m$.
          The third to fifth rows show the critical Wald-statistics at three significance levels
          according to a bootstrap approach with 2000 replications.
          The last row presents the bootstrap $p$-values.
          The optimal delay is $\hat{m}=4$, highlighted in bold face.}
  \label{TB:LinearityTest}
  \medskip
  \begin{tabular}{ccccccccccccccc}
    \hline
    $m$        &    1 & 2 & 3 & {\textbf{4}} & 5& 6 & 7 & 8 & 9 & 10 & 11 & 12 \\
    $W$        & 84.0 &100.2 &120.4 &{\textbf{127.1}} & 88.4 & 63.7 & 68.7 &102.2 &113.0 &102.9 & 87.2 &110.2\\%
    $W_{10\%}$ & 44.1 & 42.7 & 41.8 & {\textbf{43.0}} & 41.8 & 42.0 & 40.1 & 41.5 & 41.2 & 41.1 & 39.3 & 39.9\\%
    $W_{5\%}$  & 54.7 & 50.9 & 51.4 & {\textbf{52.2}} & 53.9 & 52.6 & 51.0 & 50.4 & 51.5 & 51.6 & 48.6 & 51.8\\%
    $W_{1\%}$  & 75.5 & 72.6 & 76.5 & {\textbf{72.5}} & 75.6 & 81.1 & 76.7 & 75.5 & 77.2 & 77.2 & 74.8 & 74.7\\%
    $p$-value  & $0.4\%$ & $0.1\%$ & $0.1\%$ & ${\mathbf{0.1\%}}$ & $0.4\%$ & $2.4\%$ & $1.8\%$ & $0.1\%$ & $0.1\%$ & $0.1\%$ & $0.4\%$ & $0.1\%$\\\hline
  \end{tabular}
\end{table}

The optimal value of decay $m$ can be determined exogenously, which
maximizes the value of $W$ \cite{Caner-Hansen-2001-Em}. According to
Table \ref{TB:LinearityTest}, the Wald statistic is maximized
($W=127.1$) when $m=4$. Hence, we take $\hat{m}=4$ as the optimal
decay parameter, which results in a preferred TAR model.
Accordingly, the point estimate $\hat\lambda$ of the threshold is
determined to be $-0.1418$. For the preferred specification with
$\hat{m}=4$, we report in Table \ref{TB:Parameters} the least
squares parameter estimates $\hat\theta_1$ and $\hat\theta_2$ with
standard errors.

\begin{table}[htbp]
\centering
 \caption{Least-squares estimates of parameters from the unconstrained threshold
  model with an optimal decay $\hat{m}=4$. The threshold estimate is $\hat\lambda=-0.1418$.}
 \label{TB:Parameters}
 \medskip
 \begin{tabular}{cccccccc}
    \hline\hline
    && \multicolumn{2}{@{\extracolsep\fill}c}{$Z_{t-1}<\hat\lambda$} && \multicolumn{2}{@{\extracolsep\fill}c}{$Z_{t-1}\geqslant\hat\lambda$}  \\
    \cline{3-4} \cline{6-7}
    Regressors   && Estimate & S.E.  && Estimate & S.E.   \\\hline
$y_{t-1}$          &&  0.0091 & 0.0560 && -0.0109 & 0.0197\\%
Intercept          && -0.1272 & 0.4392 &&  0.0793 & 0.1414\\%
$\Delta{y_{t-1}}$  && -1.9509 & 0.3440 &&  0.1530 & 0.0705\\%
$\Delta{y_{t-2}}$  &&  0.0504 & 0.5108 && -0.0495 & 0.0671\\%
$\Delta{y_{t-3}}$  &&  1.0005 & 0.4000 &&  0.0978 & 0.0671\\%
$\Delta{y_{t-4}}$  && -0.6906 & 0.3871 &&  0.0379 & 0.0634\\%
$\Delta{y_{t-5}}$  &&  0.5264 & 0.1916 &&  0.0852 & 0.0830\\%
$\Delta{y_{t-6}}$  && -0.1085 & 0.1564 &&  0.0415 & 0.0837\\%
$\Delta{y_{t-7}}$  &&  0.5258 & 0.1667 && -0.0564 & 0.0983\\%
$\Delta{y_{t-8}}$  && -0.6614 & 0.2292 && -0.0318 & 0.0658\\%
$\Delta{y_{t-9}}$  &&  1.4061 & 0.2966 &&  0.2429 & 0.0626\\%
$\Delta{y_{t-10}}$ &&  1.1534 & 0.3654 && -0.0484 & 0.0656\\%
$\Delta{y_{t-11}}$ && -0.5198 & 0.3796 &&  0.2115 & 0.0634\\%
$\Delta{y_{t-12}}$ &&  0.8813 & 0.2942 && -0.0290 & 0.0668\\%
\hline\hline
\end{tabular}
\end{table}

\begin{figure}[htb]
\centering
\includegraphics[width=8.5cm]{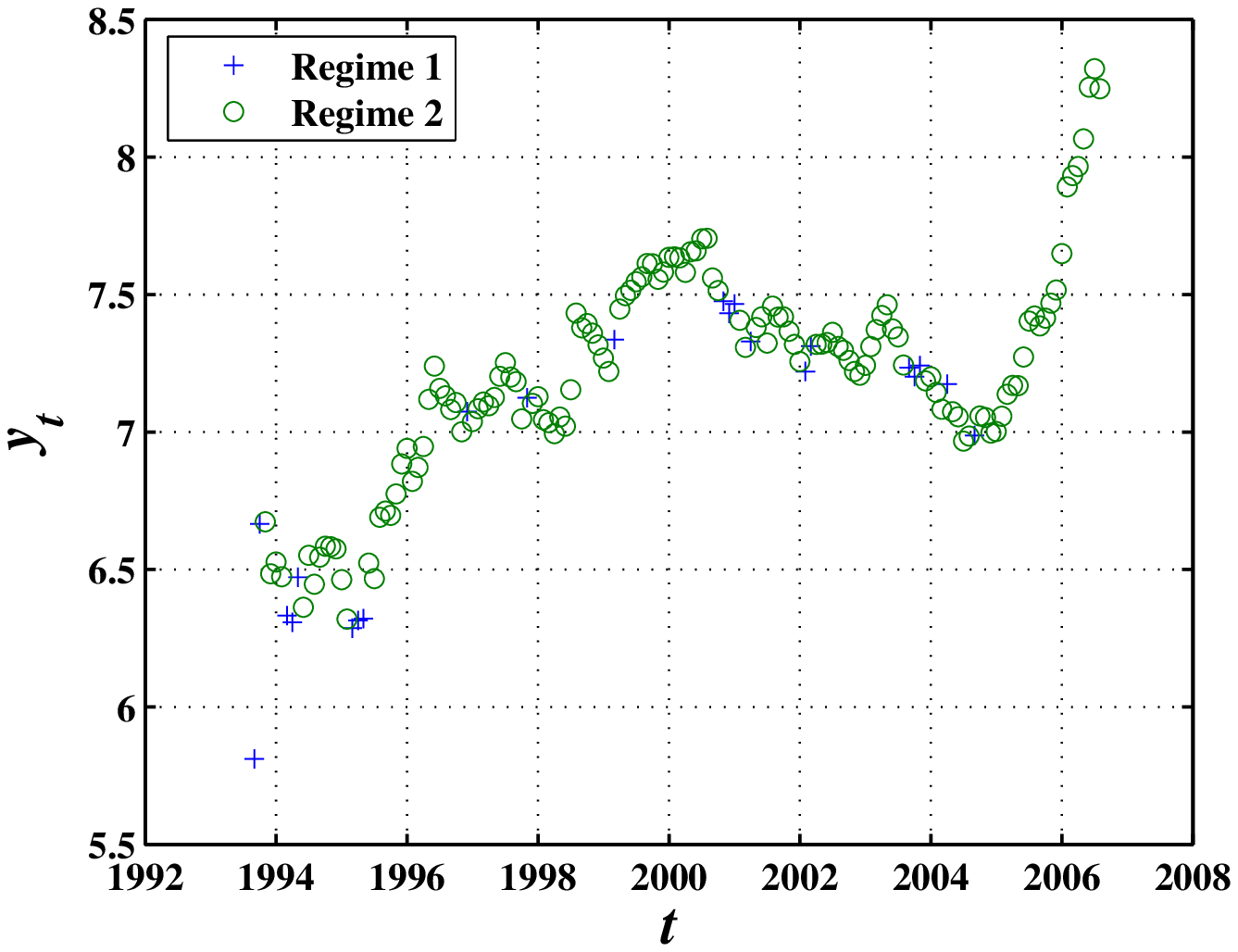}
\caption{\label{Fig:TAR:SSEC} Monthly data of the SSEC index
classified by threshold regime. The first regime (``Regime 1'' in
the legend) constitutes observations with the cumulative price drop
greater than $14.18\%$ in the last four months.}
\end{figure}

The TAR model identifies two regimes depending on whether the
variable $Z_{t}=y_{t}-y_{t-4}$ lies above or below the threshold
$\hat\lambda=-0.1418$. The first regime is when $Z_{t} < -0.1418$,
which occurs when the SSEC index has fallen cumulatively more than
$14.18\%$ in the last four months. About 16.4\% of the observations
fall into this first regime. The second regime is for $Z_{t}
\geqslant -0.1418$, which constitutes all those observations that
occur when the $m$-month price variation is no less than $-0.1418$.
Approximately $83.6\%$ of the observations belong to the second
regime. Figure \ref{Fig:TAR:SSEC} shows the estimated division of
the SSEC index into two regimes.

\subsection{Threshold unit root tests}

We examine the unit root properties of the SSEC index that possesses
significant threshold effect. We first compute the one-sided and
two-sided threshold unit root test statistics $R_1$ and $R_2$
together with the bootstrap critical values at three significance
levels $10\%$, $5\%$ and $1\%$ and $p$-values for each delay
parameter $m$, ranging from 1 to 12. The critical Wald-statistics at
three significance levels as well as the $p$-values are calculated
according to a bootstrap approach with 2000 replications. The
results are reported in Table \ref{TB:R1R2}. The one-sided Wald
tests in the left panel of Table \ref{TB:R1R2} show that the
statistic $R_1$ is less than $W_{10\%}$. The situation is similar
for the two-sided Wald tests presented in the right panel of Table
\ref{TB:R1R2}. In summary, for all $m$, both $R_{1}$ and $R_{2}$ are
less than the critical value at the $10\%$ level of significance.
These results suggest that the null hypothesis of the presence of a
unit root in the monthly SSEC index cannot be rejected at the $10\%$
level of significance.

\begin{table}[htbp]
\centering
 \caption{One-sided (left panel) and two-sided (right panel) Wald tests for threshold unit roots in the
          monthly SSEC index for different lags $m$.
          The optimal delay is $\hat{m}=4$, highlighted in bold face.}
 \label{TB:R1R2}
 \medskip
 \begin{tabular}{cccccccccccccccccc}
 \hline\hline
& \multicolumn{5}{@{\extracolsep\fill}c}{One-sided Wald test, $R_1$}
&&
\multicolumn{5}{@{\extracolsep\fill}c}{Two-sided Wald test, $R_2$}  \\ %
\cline{2-6} \cline{8-12}
$m$ & $R_1$ & $W_{10\%}$ & $W_{5\%}$ & $W_{1\%}$ & $p$-value && $R_2$ & $W_{10\%}$ & $W_{5\%}$ & $W_{1\%}$ & $p$-value \\%
\hline
1 &  0.9 & 11.2 & 14.6 & 23.9 & 0.838 &&  4.2 & 11.5 & 15.0 & 23.9 & 0.511 \\%
2 &  1.1 & 11.2 & 14.1 & 22.8 & 0.819 &&  8.3 & 11.6 & 14.5 & 23.0 & 0.216 \\%
3 &  4.5 & 11.9 & 15.6 & 27.2 & 0.446 &&  4.5 & 12.6 & 16.4 & 27.8 & 0.496 \\%
{\textbf{4}} &  {\textbf{0.3}} & {\textbf{11.7}} & {\textbf{15.4}} & {\textbf{25.5}} & {\textbf{0.922}} &&  {\textbf{0.3}} & {\textbf{12.1}} & {\textbf{15.6}} & {\textbf{25.7}} & {\textbf{0.975}} \\%
5 &  0.0 & 12.2 & 16.6 & 31.5 & 0.973 &&  0.0 & 12.5 & 17.5 & 31.5 & 0.997 \\%
6 &  1.0 & 12.0 & 16.4 & 28.0 & 0.845 &&  1.0 & 12.5 & 16.5 & 28.9 & 0.901 \\%
7 &  3.4 & 12.5 & 16.4 & 30.2 & 0.587 &&  3.4 & 12.9 & 16.7 & 30.5 & 0.647 \\%
8 &  0.1 & 13.1 & 17.2 & 28.2 & 0.954 &&  0.2 & 13.5 & 17.6 & 28.6 & 0.978 \\%
9 &  1.4 & 12.9 & 17.4 & 33.3 & 0.805 &&  1.4 & 13.5 & 18.0 & 33.9 & 0.863 \\%
10 &  1.0 & 13.0 & 17.6 & 32.0 & 0.857 &&  1.7 & 13.5 & 18.0 & 32.5 & 0.845 \\%
11 &  0.4 & 13.8 & 18.1 & 35.4 & 0.919 &&  0.7 & 14.4 & 19.1 & 37.4 & 0.935 \\%
12 &  0.4 & 14.7 & 19.7 & 42.6 & 0.911 &&  1.9 & 15.0 & 19.9 & 42.6 & 0.815 \\%
\hline\hline
    \end{tabular}
\end{table}

Although both tests $R_1$ and $R_2$ cannot reject the unit root
hypothesis, they are not able to discriminate between the full unit
root case in both regimes and the partial unit root case in one
regime. We thus test the partial unit root in the monthly SSEC index
by calculating the individual $t$ statistics, $t_1$ and $t_2$. The
results are reported in Table \ref{TB:t1t2}. The critical
Wald-statistics at three significance levels as well as the
$p$-values are calculated according to a bootstrap approach with
2000 replications. We find that, for all $m$, both $t_1$ and $t_2$
are less than the critical value at the $10\%$ significance level.
Hence, we are again unable to reject the unit root null hypothesis
in both regimes of the monthly SSEC index.

\begin{table}[htbp]
\centering
    \caption{Wald tests for threshold unit roots in the
          two regimes of the monthly SSEC index for different lags $m$.
          The optimal delay is $\hat{m}=4$, highlighted in bold face.}
    \medskip
    \label{TB:t1t2}
    \begin{tabular}{cccccccccccccccccc}
    \hline\hline
    & \multicolumn{5}{@{\extracolsep\fill}c}{$t_1$} && \multicolumn{5}{@{\extracolsep\fill}c}{$t_2$}  \\
    \cline{2-6} \cline{8-12}
 $m$ & $t$-stat &  $t_{10\%}$ & $t_{5\%}$ & $t_{1\%}$
       & $p$-value & & $t$-stat & $t_{10\%}$ & $t_{5\%}$ & $t_{1\%}$
       & $p$-value \\\hline
1 & -1.8 &  2.6 &  3.0 &  4.3 & 0.977 &&  0.9 &  2.7 &  3.2 &  4.2 & 0.539 \\%
2 & -2.7 &  2.6 &  3.0 &  4.1 & 0.996 &&  1.0 &  2.7 &  3.2 &  4.2 & 0.525 \\%
3 &  2.0 &  2.6 &  3.1 &  4.4 & 0.207 &&  0.7 &  2.8 &  3.4 &  4.6 & 0.628 \\%
{\textbf{4}} & {\textbf{-0.2}} &  {\textbf{2.5}} &  {\textbf{3.0}} &  {\textbf{4.1}} & {\textbf{0.827}} &&  {\textbf{0.6}} &  {\textbf{2.7}} &  {\textbf{3.3}} &  {\textbf{4.7}} & {\textbf{0.673}} \\%
5 & -0.2 &  2.6 &  3.1 &  4.5 & 0.826 && -0.0 &  2.8 &  3.5 &  5.1 & 0.818 \\%
6 &  1.0 &  2.6 &  3.2 &  4.2 & 0.522 &&  0.3 &  2.9 &  3.5 &  4.9 & 0.737 \\%
7 &  1.8 &  2.6 &  3.1 &  4.1 & 0.277 &&  0.0 &  2.9 &  3.5 &  5.1 & 0.812 \\%
8 & -0.3 &  2.7 &  3.2 &  4.4 & 0.871 &&  0.3 &  2.9 &  3.6 &  4.9 & 0.724 \\%
9 &  0.5 &  2.7 &  3.2 &  4.3 & 0.708 &&  1.1 &  3.0 &  3.6 &  5.6 & 0.517 \\%
10 & -0.8 &  2.6 &  3.1 &  4.2 & 0.921 &&  1.0 &  3.0 &  3.6 &  5.4 & 0.554 \\%
11 & -0.6 &  2.8 &  3.3 &  4.5 & 0.893 &&  0.6 &  3.0 &  3.7 &  5.6 & 0.658 \\%
12 & -1.2 &  2.8 &  3.4 &  4.6 & 0.952 &&  0.7 &  3.1 &  3.8 &  6.1 & 0.653 \\%
\hline\hline
    \end{tabular}
\end{table}

It is noteworthy that same conclusions are reached when we use in
the above statistical test the asymptotic $p$-valued tabulated by
Caner and Hansen \cite{Caner-Hansen-2001-Em}.

\section{Concluding remarks}
\label{s1:conclusion}

In summary, we have adopted the econometric approach of threshold
autoregression (TAR) with a unit root developed by Caner and Hansen
\cite{Caner-Hansen-2001-Em} to analyze the monthly data of the
Shanghai Stock Exchange Composite index. The SSEC index is found to
have a threshold effect of $\hat\lambda=-0.1418$ with strong
evidence. In addition, both regimes with the index variation below
or above the threshold have significant unit roots, so does the
whole time series. Our results indicate that the Shanghai stock
market exhibits nonlinear behaviors with a unit root.

An important question arises asking what we can learn further from
the fact that the stock market is nonlinear with a threshold. The
presence of a threshold $\hat\lambda=-0.1418$ means that the market
behaves differently when it falls more than $14.18\%$ in four
months. This threshold effect has direct connection with the concept
of large drawdowns in the sense of coarse graining in time for the
former and price variation for the latter, which are usually
outliers
\cite{Johansen-Sornette-1998-EPJB,Johansen-Sornette-2001-JR}. By
scanning different time scales, one might be able to provide
evidence for such a connection.

Another closely relevant issue is the definition of crashes. A
consensus is still lack. A quite feasible and unambiguous definition
is based on large drawdowns \cite{Sornette-2003,Sornette-2003-PR}. A
systematic investigation shows that more than $50\%$ of the crashes
identified are endogenously triggered that have evident preceding
log-periodic power-law (LPPL) patterns
\cite{Johansen-Sornette-2005,Johansen-2003-PA}. An alternative
option is to seek for large price drops within different time
windows \cite{Mishkin-White-2002-NBER}. These two methods identify
partially overlapping examples of crashes. It is thus interesting to
explore the possibility of having a new definition of crashes by
developing a multiscale TAR approach. The idea of multiscale
analysis is also related to the LPPL investigation of large
financial variations \cite{Sornette-Zhou-2006-IJF}. It is however
beyond the scope of the current work.

Similar to the parity pair of drawdown and drawup, it is natural to
think of the presence of a positive threshold. This needs a
three-regime threshold autoregression method with two thresholds. If
there do exist two thresholds (positive and negative) in the stock
market behavior, one can expect that the two-regime threshold
autoregression method will result in a negative threshold in some
cases as in the monthly SSEC index and a positive threshold in other
cases. This calls for further studies.

\bigskip
{\textbf{Acknowledgments:}}

We are grateful to professor B. Hansen for providing the Matlab
codes. This work was partly supported by the National Natural
Science Foundation of China (Grant No. 70501011), the Fok Ying Tong
Education Foundation (Grant No. 101086), and the Shanghai
Rising-Star Program (No. 06QA14015).

\bibliography{E:/papers/auxiliary/Bibliography}

\end{document}